\documentclass[final,3p,times]{elsarticle}
\usepackage{amsmath,amssymb,bm,color,graphicx,subfig,url,hyperref} 

\DeclareMathOperator{\sgn}{sgn}
\DeclareMathOperator{\Ei}{Ei}
\newcommand{\DD}[2]{\frac{\partial #1}{\partial #2}}
\newcommand{\DDD}[2]{\frac{\partial^2 #1}{\partial #2^{2}}}
\newcommand{\rd}{\mathrm{d}}

\journal{Wave Motion}

\begin{document}

\begin{frontmatter}

\title{Pseudolocalized Three-Dimensional Solitary Waves as Quasi-Particles}

\author[cic]{C.~I.~Christov\fnref{fn1}}
\ead[url]{http://christov.metacontinuum.com}
\address[cic]{Department of Mathematics, University of Louisiana at Lafayette, Lafayette, LA 70504, USA}

\fntext[fn1]{Prof.~C.~I.~Christov passed away after the completion of this manuscript (see preprint \href{http://arxiv.org/abs/1203.3779}{\tt arXiv:1203.3779}) but prior to its submission for publication. As a result, this submission is made by Prof.\ Ivan C.\ Christov (E-mail: \url{christov@purdue.edu}, \url{christov@alum.mit.edu}). All further correspondence can be addressed to him.}


\begin{abstract}
A higher-order dispersive equation is introduced as a candidate for the governing equation of a field theory. A new class of solutions of the  three-dimensional field equation are considered, which are not localized functions in the sense of the integrability of the square of the profile over an infinite domain. For this new class of solutions, the gradient and/or the Hessian/Laplacian  are square integrable. In the linear limiting case, an analytical expression for the pseudolocalized solution is found and the method of variational approximation is applied to find the dynamics of the centers of the quasi-particles (QPs) corresponding to  these solutions. A discrete Lagrangian can be derived due to the localization of the gradient and the Laplacian of the profile. The equations of motion of the QPs are derived from the discrete Lagrangian. The pseudomass (``wave mass") of a QP is defined as well as the potential of interaction. The most important trait of the new QPs is that at large  distances, the  force of attraction is proportional to the inverse square of the distance between the QPs. This can be considered analogous to the gravitational force in classical mechanics.
\end{abstract}


\end{frontmatter}

\section{Introduction}

The first half of the 20th century saw the formation of the current paradigm in physics, in which particles and fields are the main distinct forms of matter. Then naturally arose the question of the interconnection between these two facets of the physical reality. In the early 1960s, Skyrme \cite{Skyrme61} came up with the idea that a localized solution of a given field equation can be considered as a particle. Because of the formidable mathematical difficulties in multiple spatial dimensions, the new idea was demonstrated in 1D for the case of the $sine$-Gordon equation  ($s$GE), for which an analytical solution was found in \cite{Perring_Skyrme} for a profile composed by the superposition of two localized shapes. The shapes interacted (scattered) in a very similar fashion as two particles would do if they collided with each other.  In seemingly unrelated research, Fermi and coworkers \cite{FPU} 
discovered numerically that a virtually random initial condition for the difference equations modeling an atomic lattice tends, in the long term, to organize into a chain of localized shapes.  Observing that the Korteweg--de Vries equation (KdVE) is the limit of the difference equations modeling the lattice, Zabusky and Kruskal \cite{Zabu65} performed similar numerical experiments for KdVE and confirmed the tendency of the initially ``thermalized" modes to  organize into localized waves that interact as particles: the term \emph{soliton} was introduced for this kind of wave.

Nowadays, the subject of solitons enjoys considerable attention. The surveys \cite{Makhankov,KivMalo,KivMalo2} give a good perspective of the wide scope of  soliton research at the end of the 1970s. Applications to dynamics of atomic latices are summarized in \cite{KevreWein}. In \cite{Maugin_MRC} an excellent updated survey of the application of the soliton concept in elasticity can be found. Several general surveys are also available \cite{ScottChuMclaugh,Degasperis,AblowitzClarkson}. The list of the equations for which soliton solutions are sought is ever expanding (see, e.g., \cite{Yomba}).

If an initial condition is composed of the superposition of two localized waves, the evolution of this wave system results in the virtual recovery of the shapes of the two initial localized waves after their collision but in shifted relative positions after leaving the site of interaction (see, e.g., \cite{AblowitzClarkson,BulloughCaudrey,ScottChuMclaugh,Doddetal82}). The collision property is what justifies calling these localized waves ``quasi-particles" (QPs). A  QP solution of a fully integrable system is a soliton in the strict sense. For the case of a non-fully-integrable system, the term QP is safer, but many researchers use the term ``soliton" in a broader sense that includes also non-fully-integrable cases. When the physical system is described by equation(s) that conserve the energy and momentum, then the same conservation properties are inherited by the QPs, as would be necessary for actual subatomic particles.

What would be called a ``two-soliton" solution in the late 1960s, was given in 1962 by Perring and Skyrme \cite{Perring_Skyrme} for the $s$GE, and the resemblance between the two-soliton solution and a specially constructed superposition of one-solitons could be seen in the cited paper. In \cite{Skyrme61}, the localized solutions were qualitatively related to mesons assuming that the meson field  is governed by the $s$GE.  What is more important is that Skyrme introduced the fundamental notion that the localized (in an appropriate sense) solutions can be considered as particles of the field described by the particular equation under consideration. Using the techniques for finding analytically the two-soliton solution, an important analytical result was obtained in \cite{Yoneyama,BowtellStuart} by extracting the positions of the centers of the   individual solitons from the actual two-soliton profile. Thus the trajectories of Skyrme's ``particle" (the QP in the modern terminology) were found and shown to bend during the interaction. The last result completes, in a sense, Skyrme's proposal that the dynamics of the localized solution of the respective field equation  can describe subatomic particle interactions.  The impressive analytical success here can be clearly attributed to the full integrability of $s$GE and to the 1D nature of the solution. This line of research is currently actively pursued with the goal of creating a kind of ``multi-body" formalism based on the connection to the field equations \cite{Babelon}.  Thus, Skyrme's idea to establish the relation between the dynamics of soliton solutions of a field equation  and subatomic particles proved to be a fruitful paradigm (see,  e.g., \cite{Manton}). 

A similar approach has been developed for the integrable Camassa--Holm equation \cite{CamassaHolm}, making it possible to reduce the interaction dynamics of the so-called ``peakons'' to the solution of a system of Hamiltonian ordinary differential equations (ODEs) (see also \cite{Cotter_waltz} for an extension to cross-coupled Camassa--Holm equations). Yet another approach to the reduction of the dynamics of interacting solitons to a system of ODEs governing just a few parameters is available for the integrable nonlinear Schr\"odinger (NLS) equation, namely the adiabatic perturbation approach \cite{Gerdjikov}. 

When the field equation is not fully integrable, the way to obtain information about the QP behavior of the solution is to make an ansatz consisting of the linear superposition of two  one-soliton solutions with a priori unknown trajectories, assuming that the two-soliton solution can be fairly well approximated by the latter. Then, one derives a ``discrete" Lagrangian with the trajectories as ``generalized coordinates'' and solves the Newtonian-like governing equations for the point dynamics of the centers of the individual solitons. This was proposed in \cite{Matsuda,Sugiyama} and applied to the (non-integrable) nonlinear Klein--Gordon equation (KGE).  The trajectory function for the center of a QP was called a ``collective coordinate" in \cite{Sugiyama} (see, also, \cite{Campbell83}) and ``collective variable'' (CV) in \cite{FerguWillis}. Nowadays, this approach is known as \emph{variational approximation} (VA). The success of the simplest approach with one CV inspired the introduction of more sophisticated VA models, some involving two (or more) CVs. An important step in this direction was made in \cite{Rice}, where the second CV is the ``width" of the QP. Such a choice allows one to recover, as in \cite{Rice,Bergman}, the relativistic dynamics of a single QP in the case of KGE (see \cite{Willis88} for multiple CVs). Unfortunately, the generalization of the 2-CV method to 2-QP interaction is not trivial. For this reason the 2-CV approach was tested in \cite{FerguWillis} for the $s$GE case. The integrals for the potential of interaction are unsurmountable without some simplifying assumptions (see \cite{SG_AIMS}). 

Despite this shortcoming, the VA is clearly a very promising way of establishing the wave-particle dualism for QPs, provided that some more pertinent models for the field equation are used.  We would like to mention here that VA is one of the specific techniques to find an approximate solution of the field equations. The approach of reducing the system with distributed parameters to a discrete system amounts, in general, to a \emph{coarse grain description}  (CGD) \cite{IvanDad_sG}. From this point of view the VA is the most  physically consistent way to create a CGD of a continuous system because it retains the main conservation laws related to the latter.

Before proceeding further, it is important to mention that recently, another avenue for further development of the VA has been opened. The full two-soliton solution shows a significant deformation of the wave profile when the two main solitons are close to each other. Such a deformation is clearly not contained in the mere superposition of two one-soliton solutions. There is some utility to try to represent this deformation via the evolution of the widths of the QPs (as in the previous paragraph), but there is also another very practical approach: to consider what is left from the two-soliton solution when the one-soliton solutions are subtracted as a new particle that is briefly ``born'' at the site of interaction and promptly ``dies'' after the separation of the main lumps (the so-called ``ghost particle" \cite{Nguyen}). This idea was elaborated in \cite{BenKasmYoung} and offers a very pertinent perspective on the world of elementary particle where the birth, death, and transmutation of the particles are incessant processes.

Apart from $s$GE and KGE, there are many different classes of dispersive nonlinear equations that possess localized solutions and for which the VA proved to be a very effective tool for obtaining approximate solutions. 
The NLS equation has arisen in modeling the propagation of light along optical fibers \cite{Agrawal}. Because of the nonlinearity, localized envelopes can appear on the carrier's optical frequency, which behave like QPs. The phenomenology is much richer for NLS, and the solitons for the envelope can be ``dark'' or ``bright'' \cite{KivMalo,KivMalo2}.  The bright solitons are sometimes called ``light bullets'' \cite{Silberberg,EdmundEnns} and play a very important role in discretizing (digitizing) the information carried through the fiber. The VA was first applied to bright NLS solitons in \cite{Anderson,Karpman}, and, since then, NLS and vector NLS are the most important examples of the concept of QP. An excellent survey on this subject can be found in \cite{KivMalo,KivMalo2}. 

The concept of CGD of a field as an ensemble of QPs is clearly an important step in understanding the nature of wave-particle duality. Yet, there are essential differences between the currently known QPs and the real particles. The most conspicuous difference is that the QPs in 1D actually pass through each other because when the force is positive it remains positive up to the moment of collision, and then changes its sign.  It is interesting to note that the 1D space is so restrictive that even the two-soliton solution has a form suggesting that the individual solitons passed through each other rather than scatter (see the in-depth discussion in \cite{FerguWillis}). It turns out \cite{SG_AIMS} that only VA can shed light on what is happening, in the sense that the QPs scatter, but in this process, because of the restricted freedom in 1D, the scattered QP acquires the amplitude (pseudomass) of the other one, and the composite wave profile appears as if the QPs passed  through each other.  This effect was discovered in \cite{MillerChri} and called ``bouncing of mass" of the solitons for the case of two coupled KdVE. The repulsion or attraction of the QPs depends  on the interaction potential defined by the asymptotic behavior of their tails (see \cite{Khawaja}, among others).  The second difference between the QPs in 1D and the subatomic particles is that the attractive force decays exponentially with the distance between the QPs, while in reality, the attraction is given by Newton's law of gravitation, for which the force decays with the inverse square of the distance between the centers of the particles. This makes it important to investigate QPs in dimensions higher than one, in hopes of finding a potential of interaction that is more realistic. For this purpose any of the known generalized wave equations containing dispersion should be a legitimate basic model. 

The first 3D QPs were reported  for the NLS  in \cite{EdmundEnns,Silberberg}, where the profiles of the light bullets are found numerically for the stationary QPs, and, then, the VA is used to investigate their interaction. It is well known that the NLS is  very fertile ground for experimenting with localized shapes because of the envelope nature of the latter. (In the 1980s,  the interactions of localized 2D solutions, termed ``dromions'' \cite{Fokas1990}, to the Davey--Stewartson equation, an integrable analogue of NLS in 2D, were analyzed \cite{Boiti1988}.) In the traditional field theories, such as $s$GE and KGE, it is not straightforward to generalize QPs to higher dimensions because a solution of the type of the classical kink cannot be found in more than one dimension. Hence the simple notion of a lump of deformation of the field being a QP cannot be extended into 3D. Recently, oscillating localized 3D structures were found for the $s$GE in \cite{RosenauKashdan1,RosenauKashdan2} and called \emph{compactons}. Forcing the solution of KGE to oscillate in time leads to the same equation for the 3D shape as in the above mentioned NLS works. Finally, another avenue of exploring the dynamics of 3D QPs is through the so-called EPDiff (Euler--Poincar\'e) equation \cite{HolmStaley}, for which certain singular solutions termed ``diffeons'' \cite{Holm2013} represent 3D generalizations of the peakons of the Camassa--Holm equation \cite{CamassaHolm}.

Despite their 3D nature, the localized solutions mentioned in the previous paragraph exhibit the same exponentially decaying tails (actually, slightly super-exponential), and the respective force acting between them will decay exponentially with the increase of the distance between the QPs.

The purpose of the present paper is to explore a new field equation and a new concept of localization of the solution with the aim of bringing some of the properties of QPs closer to the observed behavior of the objects in  particle physics.

\section{The Thin Shell as a Model of a Field}

As already stated, in this paper, we are exploring the possibility of using a novel nonlinear equation to describe a field. In doing so, we would like to stay as close as possible to the spirit of quantum mechanics, i.e., as close as possible to the Schr\"odinger equation of wave mechanics, which has proven its relevance to the subject. As observed by Schr\"odinger himself \cite{Schroe1} and elaborated on in different papers of the present author (see \cite{NonProbab} and the literature cited therein), if written for the real or imaginary part of the wave function, Schr\"odinger's equation belongs to the class of Euler--Bernoulli equations \cite{Segel}  describing the flexural deformations of a thin plate. While, the Euler--Bernoulli equation is the linear part of any shell/plate equation, the nonlinear terms depend on the assumptions about the nature of the deformations. In the theory of prolate deformations of shallow shells/plates, the von K\'arm\'an \cite{Karman} equations are the current standard \cite{Ciarlet_vonKaman}. In the other extreme case of very steep deformations, the shell equation is akin to the Boussinesq equation, but the nonlinearity is different (see \cite{NonProbab} and the literature cited therein). Since, the elementary particles have very small dimensions, the QPs that are supposed to describe them may exhibit, even for small elevations,  deformations of order of unity and very large curvature (exactly the opposite  to what is considered as a prolate deformation). We choose the model from  \cite{NonProbab}, which is essentially a non-prolate model, as our field equation, namely
\begin{equation}
u_{tt} = \Delta u - \alpha_3 (\Delta u)^3 - \beta^2\Delta^2 u, \label{eq:nonprobab}
\end{equation}
where $\beta^2$ is the dispersion coefficient due to the stiffness of the shell (always positive) taken as the square of some parameter $\beta$ for convenience.  The coefficient $\alpha_3$ gives the relative importance of the nonlinear term. The membrane tension (the coefficient of the Laplacian term) is taken to be equal to unity because, for the case of localized solutions over an infinite interval (no boundary conditions at specific points), it can be scaled out.

\begin{figure}[ht]
\centerline{\includegraphics[width=0.5\textwidth]{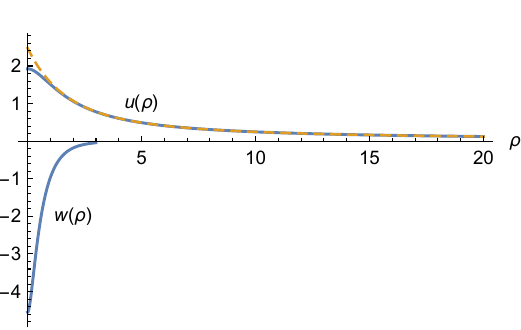}}
\caption{The pseudolocalized solution (upper part) and the localized ``spherical $sech$" (lower part) for $\beta=1$. Solid lines: numerically computed profile; dashed line: best fit solution to the linearized problem given by $2.47546(1-e^{-\rho})/\rho$.}\label{fig:NonProbab}
\end{figure}

Now, the spherically-symmetric, stationary solution satisfies
\begin{subequations}
\begin{align}
\beta^2\frac{1}{\rho^2}\frac{\rd}{\rd \rho}\left(\rho^2\frac{\rd w}{\rd \rho}\right) - w +w^3 &= 0, \label{eq:curvature} \\
 \frac{1}{\rho^2}\frac{\rd}{\rd \rho}\left(\rho^2\frac{\rd u}{\rd \rho}\right) &= w(\rho),  \label{eq:wave_funct}
\end{align}
\end{subequations}
where $-w$ is the curvature of the flexural deformation. Note that Eq.~\eqref{eq:curvature} is the same as the one solved numerically in \cite{EdmundEnns} and \cite{RosenauKashdan2}. Because of the absence of biharmonic terms in these works, the result for $w$ is the final result. For the time being, no analytical solution is available for this equation, and under the boundary condition $w(\rho) \rightarrow 0$ for $\rho\rightarrow \infty$ it has a solution that decays at infinity as $e^{-\rho}/\rho$. This solution is presented in the lower portion of Fig.~\ref{fig:NonProbab}. It corresponds to the $sech$ solution in 1D and can be termed 
the ``spherical $sech$."

After obtaining the solution for $w$, the function $u$, which can be called ``the wave function,''  can be found by quadratures, namely
\begin{equation}
u(\rho) = \int_0^\rho \left( \frac{1}{\rho_1^2} \int_0^{\rho_1} \rho_2^2 w(\rho_2) \mathrm{d} \rho_2 \right) \mathrm{d} \rho_1 + \frac{C_1}{\rho} + C_2. \label{eq:integrate_zeta}
\end{equation}
Here, one has to take $C_2=0$ in order that $u$ decays at infinity. Respectively, $C_1=0$ ensures that there is no singularity at the origin of the coordinate system. As suggested in \cite{NonProbab}, one can find a best-fit approximation to the numerical solution for $w$ and then integrate it twice according to Eq.~\eqref{eq:integrate_zeta}. The result is presented in the upper portion of Fig.~\ref{fig:NonProbab}. It is seen that the decay of the profile of $u$ at infinity is as $1/\rho$, which is significantly slower than exponential decay.

The nonlinear terms are important for shaping the profile near the origin of the coordinate system (as can be seen in Fig.~\ref{fig:NonProbab}), but at larger distances Eq.~\eqref{eq:nonprobab} reduces to the following linear equation:
\begin{equation}
u_{tt} = \Delta  u - \beta^2 \Delta^2 u, \label{eq:Euler_Bernoulli}
\end{equation}
which is the celebrated Euler--Bernoulli equation. As shown in \cite{NonProbab} this equation is akin to the Schr\"odinger equation, so interrogating Eq.~\eqref{eq:Euler_Bernoulli} is a good place to start. The Lagrangian for Eq.~\eqref{eq:Euler_Bernoulli} reads
\begin{gather}
L = \iiint_V \frac{1}{2}
\left[ \left({u}_{t}\right)^2 - (\nabla u)^2 
- \beta^2 (\Delta u)^2\right] \rd^3 \bm{x},
\label{eq:Lagrange}
\end{gather}
where, in the case under consideration, the volume of integration $V$ is the entire space. Without fear of confusion, we will use the notation $-\infty$ and $\infty$ as the limits of the triple integral in what follows.

The linearized Eq.~\eqref{eq:Euler_Bernoulli} has the solution 
(see \cite{NonProbab})
\begin{equation}
\Phi(\rho) = \frac{1}{\rho}[1 - e^{-{\rho}/{\beta}}], \label{eq:spherical_symmetry}
\end{equation}
which is compared to the solution of the nonlinear equation in Fig.~\ref{fig:NonProbab}. One can see that  the full nonlinear solution does not differ quantitatively from the linear solution given by Eq.~\eqref{eq:spherical_symmetry}, save for the fact that the latter is not smooth at the origin as function of the variables $x,y,z$. It decays at infinity, but is not localized in the strict sense because the integral of its square diverges linearly at infinity with the radius of the domain of integration. However, the integral of the square of its gradient converges:
\begin{equation}
\iiint_{-\infty}^{\infty} (\nabla \Phi)^2 \rd x \rd y \rd z \equiv 
 4\pi \int_{0}^{\infty}  \left[\frac{\rd \Phi(\rho)}{\rd \rho}\right]^2 \rho^2 \rd \rho
  = 4\pi  \int_{0}^{\infty}  \left[\frac{e^{-\rho/{\beta}}}{\rho{\beta}}-\frac{1 - e^{-\rho/{\beta}}}{\rho^2} \right]^2 \rho^2 \rd \rho
    =\frac{2\pi}{\beta}. 
 \label{eq:squared_gradient}
\end{equation}
As it should be expected, if the gradient is in $L^2(\mathbb{R}^3)$,  the Laplacian belongs to the same space, in the sense that the integral of the square of the Laplacian converges too:
\begin{equation}
\iiint_{-\infty}^{\infty} [\Delta \Phi]^2 \rd x \rd y \rd z \equiv 
 4\pi \int_{0}^{\infty} \left[\frac{1}{\rho^2}\frac{\rd}{\rd \rho} \rho^2\frac{\rd \Phi(\rho)}{\rd \rho}\right]^2 \rho^2 \rd \rho
  = 4\pi  \int_{0}^{\infty} \left[-\frac{e^{-\rho/{\beta}}}{\rho\beta^2} \right]^2 \rho^2 \rd \rho  =\frac{2\pi}{\beta^3}. 
 \label{eq:squared_Laplacian}
\end{equation}
Because of the quantitative closeness of the solution of the nonlinear equation to the linear one, the statements about the convergence and/or non-convergence of the different integrals is the same for the solutions of the full and linearized equations.  Following \cite{NonProbab}, we refer to this kind of solution as the \emph{pseudolocalized} solution. The quantitative closeness to the linear  solution allows us to investigate the interaction properties of the pseudolocalized QPs by analytical means using the result in 
Eq.~\eqref{eq:spherical_symmetry}.

It is interesting to observe here that in 2D, the solution of Eq.~\eqref{eq:Euler_Bernoulli}, with radial symmetry, that does not exhibit a singularity at the origin (see \cite{Chri_pseudo_numer}) is
\begin{equation}
u^{(2)}(r)=K_0(r/{\beta}) + \ln (r/{\beta}),
\end{equation}
where $K_0$ is the modified Bessel function of the second kind of order zero. Clearly, while this solution does not have a pole at the origin, it is not a localized function because it grows logarithmically as $r \rightarrow \infty$. It is not even a pseudolocalized one in the sense of the definition of the present work because the following integral diverges
\begin{equation}
\int_0^R \left[\frac{\rd u^{(2)}(r)}{\rd r}\right]^2 r \rd r \propto \ln R + \text{non-singular terms}, \quad\text{for}\quad R\gg 1.
\end{equation}
This means that in 2D, pseudolocalized solutions of the Euler--Bernoulli equation \eqref{eq:Euler_Bernoulli} do not exist.

\section{Localized versus Pseudolocalized Solutions}

Eq.~\eqref{eq:nonprobab} belongs to the generic class of Boussinesq equations (BEs). In different models of flows of ideal fluids with free surfaces or interfaces, different BEs arise. A  generic version of Boussinesq's equation can be written as
\begin{equation}
u_{tt} = \Delta [ u - \alpha_2 u^2 -\alpha_3 u^3 - \beta^2\Delta u]. \label{eq:BE}
\end{equation}
Eq.~\eqref{eq:BE} is a generalized wave equation containing dispersion in the form of a biharmonic operator of the sought function. The dispersive effects of the biharmonic operator can be countered by the presence of a nonlinearity, and as a result a permanent wave of localized type may exist and propagate without change. This balance between nonlinearity and dispersion was first found by Boussinesq \cite{boussi72} and we shall refer to it as the ``Boussinesq paradigm.''  In different physical situations leading to the same generic class of equations, the nonlinear terms need not be the same. In fluid mechanics, the most general form of the weakly-nonlinear approximation for flows in thin fluid layers is discussed in \cite{Benney_Luke,Chri_WM}. 

Unfortunately, for the time being we are unable to find a 3D two-soliton solution for equations of the type of Eq.~\eqref{eq:BE}. Just as in 2D, multi-soliton solutions are not known, except when the equation is simplified in one of the spatial directions (such as the so-called Kadomtsev--Petviashvili equation (KPE), see the original work \cite{KadomPetvi} and general surveys \cite{BulloughCaudrey,Degasperis}; or the so-called ``Davey--Stewartson 1'' equation (DS1E), see the original work \cite{Davey101} and the derivation of solutions \cite{Boiti1988,Fokas1990}).

The findings about the 2D localized solution of the  Boussinesq equation can be summarized as follows. The profile of the  soliton has a super exponential decay at infinity, namely, $e^{-r}/\sqrt{r}$. This property is not robust, and even for very small propagation speeds, the decay at infinity changes to an algebraic one proportional to $1/r^2$. The most disappointing property is, however, the fact that the steadily propagating 2D shapes are not robust (structurally stable), and, after some time, they either dissipate or blow up. This important result was first found by means of a special new scheme in \cite{ChertChriKurgan} and confirmed quantitatively with a different numerical approximation \cite{ChriKolkVasi}. All these findings rule out the strictly localized solutions as QPs.  The non-robustness with respect to the parameter representing the phase speed of the structure tells us that in order to make the standing localized wave move, the profile behavior up to infinity has to be \emph{instantaneously} changed, which is hard to imagine as a physically realistic process.

In order to find out if pseudolocalized solutions of  Eq.~\eqref{eq:BE} can exist we take $\alpha_3=0$  and consider the spherical symmetric stationary solution $u = G(\rho)$:
\begin{equation}
\Delta_\rho \left[ G - \alpha_2 G^2 - \beta^2 \Delta_\rho G \right] = 0, 
\end{equation}
where $\Delta_\rho$ is the Laplace operator in spherical coordinates, already defined in Eq.~\eqref{eq:squared_Laplacian}. The last equation can be integrated once to obtain
\begin{equation}
 G - \alpha_2 G^2 - \beta^2  \frac{1}{\rho^2}\frac{\rd }{\rd \rho}\left[ \rho^2 \frac{\rd G}{\rd \rho}\right] =\frac{C_1}{\rho} + C_2.
\end{equation}
In order that the solution decays at infinity we set $C_2=0$. Now, setting $C_1=0$ gives the strictly localized solution, which has been discussed in previous paragraphs as not suitable for our purposes. On the other hand, $C_1\ne 0$ leads to a pseudolocalized solution. It is important to note that the linearized version of the last equation has the same solution as the one given in Eq.~\eqref{eq:spherical_symmetry} for the shell-equation \eqref{eq:Euler_Bernoulli}. Let us call the last solution a ``peakon,'' borrowing the terminology introduced in \cite{CamassaHolm} in a different physical situation.

A special numerical investigation of the role of the nonlinearity for defining the shape of the pseudolocalized  solution was recently conducted  in \cite{Chri_pseudo_numer}. The result is that  the inclusion of nonlinearity does not qualitatively change the profile given in Eq.~\eqref{eq:spherical_symmetry}. The nonlinearity can make the profile slightly steeper at the origin, but the asymptotic behavior that is essential for the interaction of two QPs remains unchanged. All this means that, even in the case of a generic Boussinesq equation, the pseudolocalized solutions exist and are valid candidates for QPs.  Actually, in this case, the solution $\Phi(\rho)$ is an even better quantitative approximation for the profile of the QP, hence it can be used as the basis for a coarse grain description of the field.

\begin{figure}[ht]
\centerline{\includegraphics[width=0.425\textwidth]{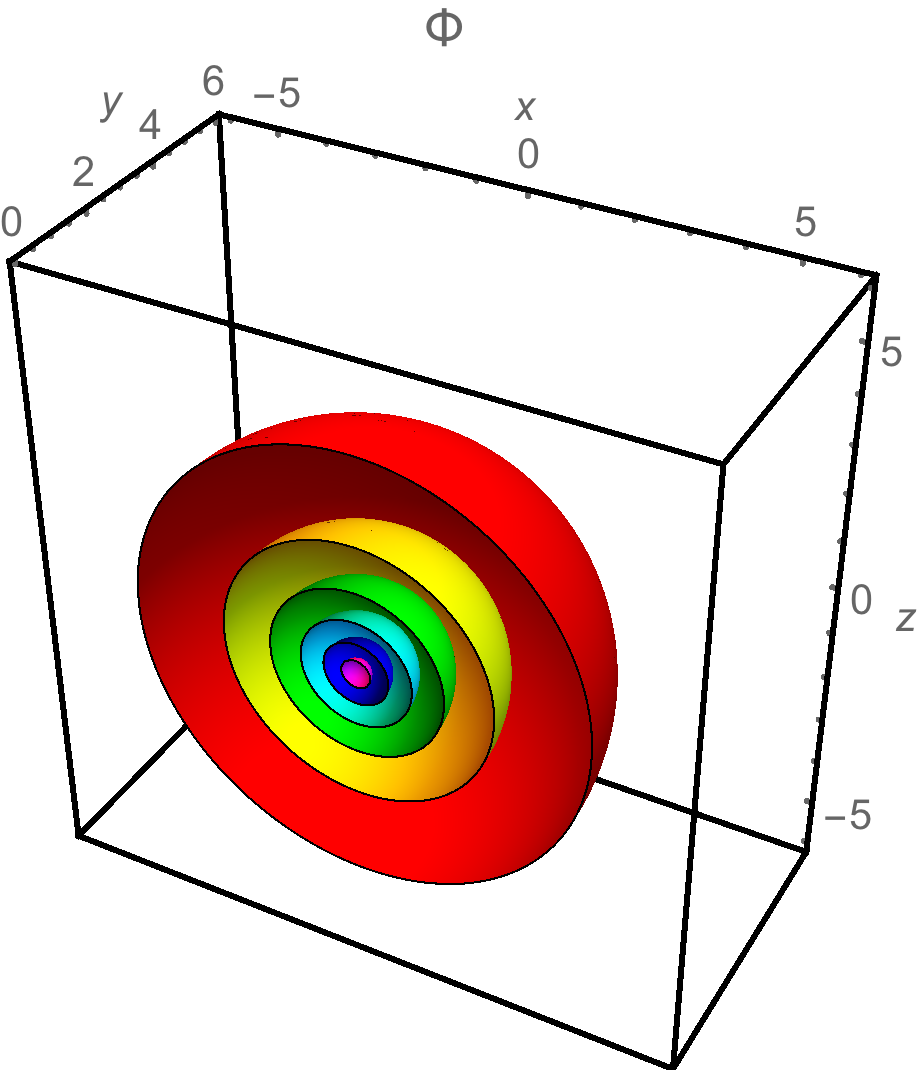}}
\caption{Resected ball representing a single radial peakon for $\beta=1$.} \label{fig:SinglePeakon} 
\end{figure}

The amplitude $\Phi(\rho)$ can be interpreted as the  elevation of the middle 3D surface of a shell (say, along a fourth dimension). To an observer confined to the middle 3D surface it would appear as some kind of density. For this reason, we use a density plot in Fig.~\ref{fig:SinglePeakon}. In order to see  the internal structure more clearly, we show only half of the ball corresponding to the main part of the peakon. Naturally, the latter is not exactly a ball because it extends to infinity but the density is smaller far from the center, so if we limit the number of contours to a certain value (sphere), the outermost contour shown is a  ball.

\section{Coordinate System for Steadily Propagating Pseudolocalized Waves}

The full three-dimensional description of the interaction of QPs of the pseudolocalized type requires significant computational resources. It is well known that, in 1D, propagating QPs experience contraction or, in some cases, elongation of the overall scale of the support. In 2D, the change of shape is more intricate (see \cite{ChriChou_asymp}), leading to relative contraction in the direction of motion, on the background of overall elongation of the scales of the solution. The first decisive simplification here will be to assume that, when propagating, the QPs do not actually change their shapes, retaining their spherical symmetry. This is a typical assumption when applying the VA.

The second important simplification is to consider QPs that interact only along the line connecting their centers. Since we have already assumed that the shape retains its spherical symmetry during the interaction, this is not really an additional restriction on the solution. Even if the QPs move with arbitrary  trajectories in the 3D configurational space, the force acting between them will depend only on the distance between the QPs, as measured along the line connecting their centers. For this reason, we restrict ourselves to the case when the QPs move only along one of the spatial directions, say the $z$-axis. Then, we can consider the radially symmetric cylindrical coordinates $r = \sqrt{x^2+y^2}$ and $z$. 
Their relation to the spherical coordinates is $\rho :=\sqrt{r^2+z^2}$ and $\theta := \arctan(r/z)$. Then,
\begin{equation}
r = \rho \sin\theta, \quad z = \rho \cos\theta.
\end{equation} 
The cylindrically symmetric Laplace operator is given by
\begin{equation}
\Delta : = \frac{1}{r} \DD{}{r} r \DD{}{r} + \DDD{}{z}. \label{eq:cyl_Laplace}
\end{equation}

\section{Coarse-Grain Description using the Field of a Single QP}

For a localized wave moving in the $z$-direction according to the law of motion $z=Z(t)$,  without changing its shape, one has
\begin{equation}
u(r,z,t) = A \Phi\left(\sqrt{r^2 + [z-Z(t)]^2}\,\right),
\end{equation}
where $A$ is its amplitude.  At this stage, we do not consider the changes of the effective support of the wave (pseudo-Lorentzian contraction or elongation) associated with the motion. In other words, the speed $\dot Z(t)$  is small in comparison with unity. Then,
\begin{equation}
u_t = A\Phi'\left(\sqrt{r^2 + [z-Z(t)]^2}\,\right) \frac{\dot Z(z-Z)}{\sqrt{r^2 + (z-Z)^2}}.\label{eq:time_detiv_1QP}
\end{equation}
Consequently, the integral of the kinetic energy density over space is:
\begin{equation}
\begin{aligned}
\frac{1}{2}\iiint_{-\infty}^{\infty} (u_t)^2  r \rd r \rd z &= 
\pi{\dot Z}^2 A^2\iint_{-\infty}^{\infty} [\Phi'(\rho)]^2 \frac{(z-Z)^2}{r^2 + (z-Z)^2}r \rd r \rd z  \\
&= \pi{\dot Z}^2A^2\int_{0}^{\infty}\int_{0}^{\pi} \big[\Phi'(\rho)\big]^2 \rho^2 \cos^2\theta \sin\theta \rd \rho \rd \theta\\ 
&= \frac{2\pi}{3}{\dot Z}^2A^2\int_{0}^{\infty} \big[\Phi'(\rho)\big]^2 \rho^2 \rd \rho \\
&= \frac{1}{2} \mathfrak{m}A^2 {\dot Z}^2, \qquad  \text{where}\quad \mathfrak{m}:= \frac{4\pi}{3} \int_{0}^{\infty}  [\Phi'(\rho)]^2 \rho^2 \rd \rho  = \frac{2\pi  }{3\beta}.
\end{aligned}
\label{eq:pseudomass}
\end{equation}
Note that, here, the relation $z-Z = \rho\cos\theta$ is used in manipulating the integrals.
In the last line of Eq.~\eqref{eq:pseudomass}, Eq.~\eqref{eq:squared_gradient} is also acknowledged.  Respectively,
\begin{equation}
\nabla u = A\Phi'\left(\sqrt{r^2 + [z-Z(t)]^2}\,\right) \frac{(r,z-Z)}{\sqrt{r^2 + (z-Z)^2}},
\end{equation}
where $(r, z-Z)$ is the vector with the respective components.
Then, it follows from Eq.~\eqref{eq:squared_gradient} that
\begin{equation}
\frac{1}{2}\iiint_{-\infty}^{\infty} (\nabla u)^2 \rd x\rd y \rd z = \frac{\pi A^2}{\beta}.
\end{equation}
In similar fashion, from Eq.~\eqref{eq:squared_Laplacian}, we obtain
\begin{equation}
\frac{\beta^2}{2}\iiint_{-\infty}^{\infty} (\Delta u)^2 \rd x \rd y \rd z    =\frac{\pi A^2}{\beta}. 
 \label{eq:potential_from_biharmonic}
\end{equation}

Now, for the discrete Lagrangian we obtain
\begin{equation}
\mathbb{L} = \frac{1}{2}\mathfrak{m}A^2\dot{Z}^2 - \frac{\pi A^2}{\beta} - \frac{\pi A^2}{\beta},
\label{eq:disc_lagr}
\end{equation}
and the Euler--Lagrange equation for the collective variable $Z(t)$ reads 
\begin{equation}
\frac{\rd}{\rd t}\left[ \mathfrak{m} \frac{\rd}{\rd t} Z\right] = 0, 
\end{equation}
which is nothing else but Newton's law of inertia. This suggests that we interpret the quantity $\mathfrak{m}$ as the mass of the QP concentrated in the center of the localized wave. Thus, we call $\mathfrak{m}$ the \emph{pseudomass} or the ``wave mass" of a QP. The ``effective'' mass $\mathfrak{m}A^2$ is then simply proportional to the square of the amplitude of the quasi-particle (see Eq.~\eqref{eq:disc_lagr}).

Since, the solutions can have either positive or negative amplitude, one can also define a quantity called \emph{pseudovolume} as the integral of the curvature:
\begin{equation}
\mathfrak{v} := -2\pi \int_0^\infty \int_0^\pi A \Delta \Phi(\rho) \rho^2 \sin\theta \rd \theta  \rd \rho = 4A\pi 
\end{equation}
This means that, due to the convention with the minus sign, in the case of positive amplitude ($A>0$), we have a positive {pseudovolume} ($\mathfrak{v}>0$). It is interesting to note that the  {pseudovolume} does not depend on $\beta$. 

The introduction of the pseudovolume allows us to better understand the notion of anti-particle. The mass of an anti-quasi-particle (anti-QP) is positive, while the volume is negative. In the response to a force, the anti-QP (negative volume) will behave as matter but when it overlaps with a QP (positive volume) they will cancel over each other.

\section{Coarse-Grain Description of the Field of Two Pseudolocalized QPs}

\begin{figure}[ht]
	\centering
	\subfloat[][distance between centers is 1.18]{\includegraphics[width=0.425\textwidth]{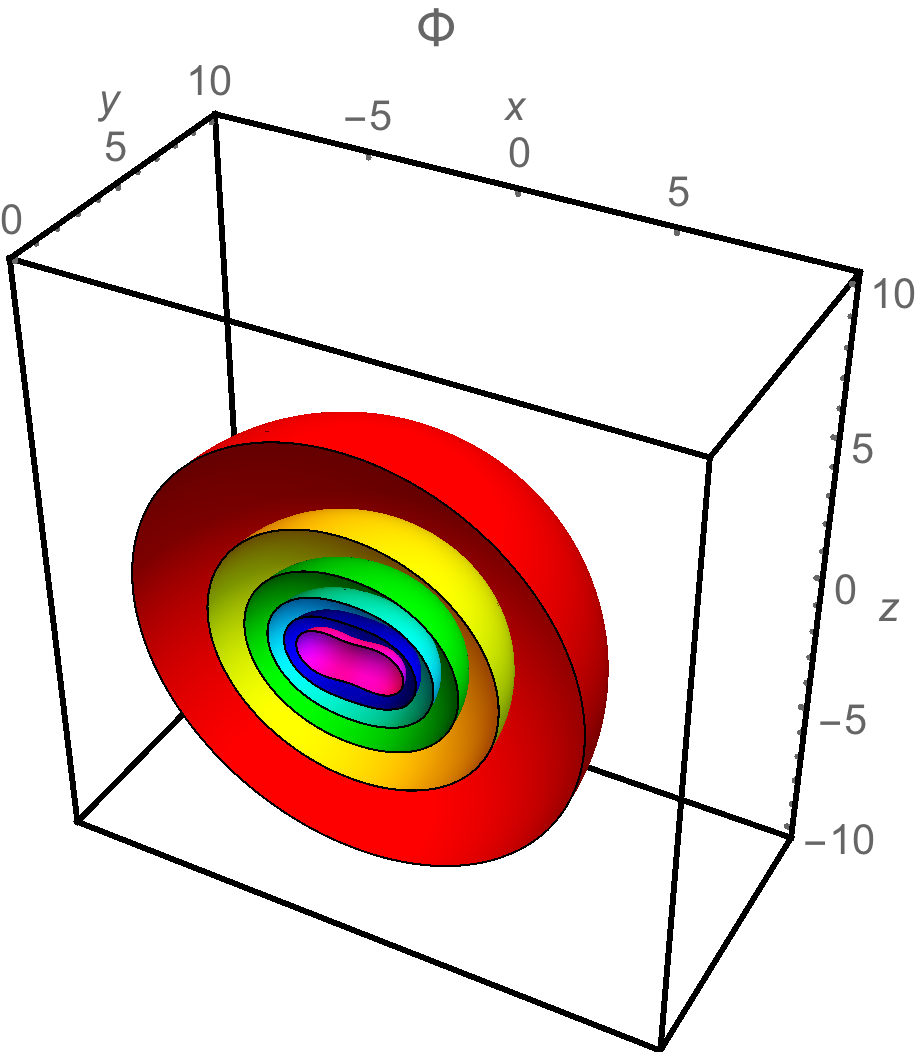}}
	\hfill
	\subfloat[][distance between centers is 4]{\includegraphics[width=0.425\textwidth]{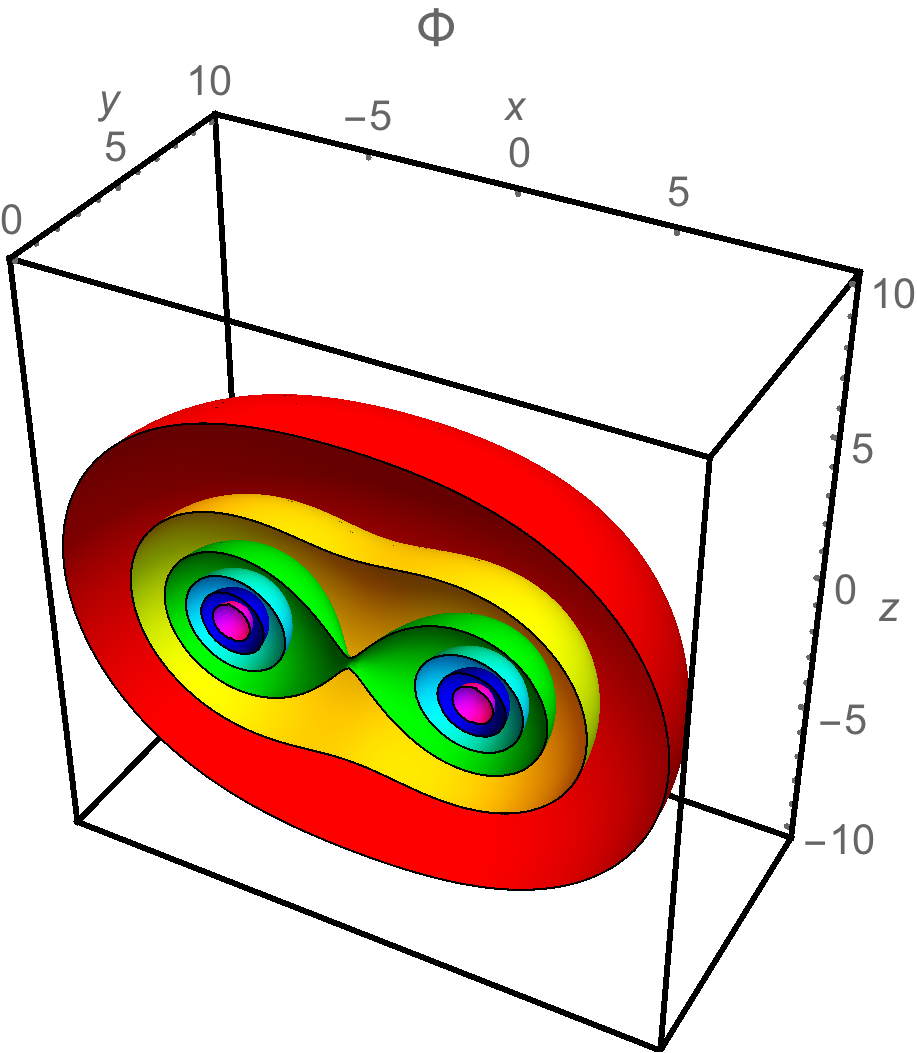}}
\caption{Two radial peakons with $A_L=A_R=1$ and two different distances between their centers (a,b); $ \beta=1$.}
\label{fig:TwoRadialPeakons}
\end{figure}

The main idea of VA is to use a ``reasonable'' approximation for the shape of the one-wave solution and then, from the continuous Lagrangian, derive the discrete Lagrangian and the laws of motion for the centers of the QPs involved. When a CGD is attempted, the shapes of the QPs are each taken to be the shape of a single stationary solitary wave.  For the case under consideration, we stipulate that
\begin{equation}
u = A_L \Phi(\rho_L) + A_R \Phi(\rho_R), \qquad \rho_k :=\sqrt{r^2+(z-Z_k)^2},\label{eq:two_QPS}
\end{equation}
where $k$ assumes the ``value'' $L$ for the left QP and $R$ for the right QP.
Here, the constraint acknowledges that the QPs are supposed to move only along the axes that connect their centers ($z$-axis in the adopted convention). The case of two identical radial peakons ($A_L=A_R=1$) with $\beta=1$ is depicted in Fig.~\ref{fig:TwoRadialPeakons} for two different distances between their centers.

In order not to obscure the main idea, we will not consider here the dependence of shape on the phase speed(s) $c_k:=\dot Z_k(t)$. This amounts to the assumption that the phase speeds are small. Such an assumption is strictly valid only at the initial stage after the QPs are allowed to move under the forces that arise due to the presence of other QPs. 

For the time derivative of the function $u$ from Eq.~\eqref{eq:two_QPS}, we find (see the one-QP case from Eq.~\eqref{eq:time_detiv_1QP}):
\begin{equation}
u_t = -A_L\Phi'(\rho_L) \frac{\dot Z_L(z-Z_L)}{\rho_L} 
- A_R\Phi'(\rho_R) \frac{\dot Z_R(z-Z_R)}{\rho_R}.\label{eq:time_deriv_2QP}
\end{equation}
In some places we suppress the explicit time argument of $Z_k$ for the sake of clarity.
Similarly, for the spatial derivatives we have
\begin{align}
u_z&= A_L \Phi'(\rho_L) \frac{(z-Z_L)}{\rho_L} 
+ A_R\Phi'(\rho_R) \frac{(z-Z_R)}{\rho_R},\label{eq:z_deriv_2QP}\\
u_r&=A_L \Phi'(\rho_L) \frac{r}{\rho_L} 
+ A_R\Phi'(\rho_R) \frac{r}{\rho_R}.\label{eq:r_deriv_2QP}
\end{align}

\subsection{Potential due to the Laplacian term}

Now, turning to the potential of interaction, we acknowledge that there are two terms in the Lagrangian, containing $\nabla u$ and $\Delta u$, respectively. For the former we have 
\begin{multline}
\frac{1}{2}\iiint_{-\infty}^{\infty} (\nabla u)^2 \rd x\rd y \rd z = 
\pi\int_{-\infty}^{\infty} \int_0^\infty [(u_r)^2+ (u_z)^2]  r \rd r \rd z 
=\pi \frac{A_L^2+A_R^2}{\beta} +  A_LA_RV_a(w),\quad\text{where}\\ 
V_a(w) := 2\pi\int_{-\infty}^{\infty} \int_0^\infty  \frac{r^2 + (z-Z_L)(z-Z_R)}{\rho_L\rho_R}   \Phi'(\rho_L)\Phi'(\rho_R) r \rd r \rd z.
\label{eq:2QP_potential_Laplace}
\end{multline}
Let us now introduce a new variable $\zeta = z -Z_L$ and the notation $w = Z_R - Z_L$.  Respectively, we keep the notation $\rho_L  = \sqrt{r^2 +\zeta^2} = \rho$ and $\rho_R = \sqrt{r^2 + (\zeta - w)^2}= \sqrt{\rho^2 - 2 \rho w \cos\theta +w^2}$. Then, the potential due to the Laplacian term is
\begin{equation}
V_a(w)  = 2\pi \int_0^{\infty} \int_0^\pi  \frac{\rho^2 - \rho w\cos\theta}{\rho\rho_R}
 \Phi'(\rho)\Phi'(\rho_R)\rho^2 \sin \theta \rd \theta \rd \rho.
\label{2QP_potential_Laplace2}
\end{equation}
After performing the integration and applying some standard manipulations (see \ref{app:laplacian}), we obtain
\begin{equation}
V_a(w) = \frac{2\pi}{\beta}e^{-{w}/{\beta }} \left\{ -\gamma + \log \left({2 \beta }/{w}\right) + \Ei\left({w}/{\beta }\right)+e^{{2 w}/{\beta }} \left[\Ei\left(-{2 w}/{\beta }\right)-\Ei\left(-{w}/{\beta }\right)\right] \right\}, 
\label{eq:2QP_potential}
\end{equation}
where $\gamma\approx 0.5772156649$ is the Euler constant and $\Ei(\xi) :=\int_{-\infty}^\xi e^t t^{-1} \rd t$ is the exponential integral. 

The result closest to the present one, regarding the interaction integrals and their computation, that can be found in the literature is given in \cite{Malomed2D3D} and elaborated in \cite{MaimitsMaloDesyat}, which is one of the few treatises on the interaction of 2D and 3D solitons. The expressions from the cited works differ from ours because they are for strictly localized QPs.

\subsection{Potential due to the biharmonic term}
In this subsection, we deal with the last term in Eq.~\eqref{eq:Lagrange}. To this end, we observe that
\begin{equation}
\Omega(\rho) := \Delta \Phi(\rho) = \frac{1}{\rho^2} \DD{}{\rho} \rho^2 \DD{\Phi}{\rho} = -\frac{e^{-\rho/\beta}}{\beta^2\rho}.
\end{equation}
Then, for the assumed profile consisting of the superposition of two QPs of the type defined in the Eq.~\eqref{eq:spherical_symmetry} we have
\begin{multline}
\frac{\beta^2}{2}\iiint_{-\infty}^\infty (\Delta u )^2  \rd x \rd y \rd z = \pi \frac{A_L^2+A_R^2}{\beta} +   A_LA_R V_b(w),\quad \text{where}\\ 
V_b(w) := 2\pi {\beta^2} \int_{-\infty}^{\infty} \int_0^\infty \Omega(\rho)\Omega(\rho_R)  r \rd r \rd z
=  \frac{2\pi}{\beta} e^{-w/\beta}.
\label{eq:2QP_potential_bihar}
\end{multline}
The last expression is obtained by using Eq.~\eqref{eq:Vb}. Clearly, the potential due to the biharmonic term is positive but not strong enough to change the main behavior of $V_a(w)$.

\subsection{Total potential and total force}

\begin{figure}[h]
	\centering
	\subfloat[][linear]{\includegraphics[width=0.49\textwidth]{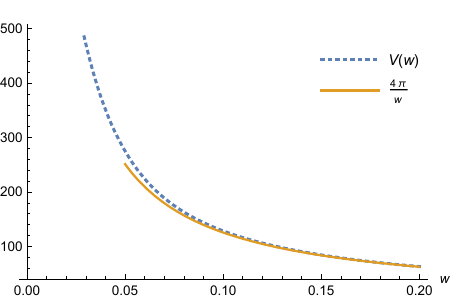}}
	\hfill
	\subfloat[][log-log]{\includegraphics[width=0.49\textwidth]{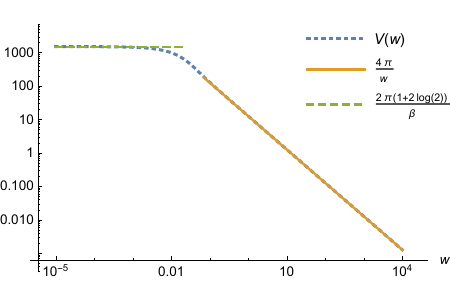}}
\caption{The full interaction potential $V(w)$ between two QPs for $\beta=0.01$.}\label{fig:potential}
\end{figure}

In order to complete the description of the dynamical properties of the system of two QPs, we need the crossmass (see the next subsection) and the total potential
\begin{equation}
V(w) := V_a(w) + V_b(w), \label{eq:total_potent}
\end{equation}
with $V_a$ and $V_b$ as they were defined in Eq.~\eqref{eq:2QP_potential} and Eq.~\eqref{eq:2QP_potential_bihar}, respectively.
In order to save space, we will not write explicitly the expression for the sum because both components are well specified above. 
The total potential is shown in Fig.~\ref{fig:potential}.
The most important feature of  $V(w)$ is the portion of the force acting at very large distances with potential $1/w$, which can be called ``gravitational" because of its one-to-one correspondence with the respective attraction law.

\begin{figure}[h]
	\centering
	\subfloat[][linear]{\includegraphics[width=0.49\textwidth]{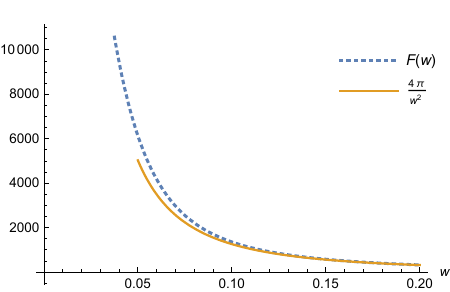}}
	\hfill
	\subfloat[][log-log]{\includegraphics[width=0.49\textwidth]{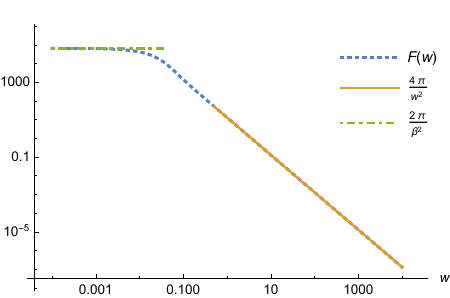}}
	\caption{The full force of interaction $F(w)$ between two QPs for $\beta=0.01$.}
\label{fig:force}
\end{figure}

The total force is
\begin{equation}
F(w) := -\frac{\rd V(w)}{\rd w} =
\frac{2\pi}{\beta^2}e^{-{w}/{\beta}} \left\{ 1-\gamma + \log \left({2 \beta }/{w}\right) + \Ei\left({w}/{\beta }\right)
- e^{{2 w}/{\beta }} \left[\Ei\left(-{2 w}/{\beta }\right)-\Ei\left(-{w}/{\beta }\right)\right] \right\}, 
\label{eq:force}
\end{equation}
and a comparison to different best fit functions/asymptotic behaviors is shown in Fig.~\ref{fig:force}. 

\subsection{Crossmass}
For the kinetic part of the energy of the system of two QPs, we derive
\begin{equation}
\frac{1}{2}\iiint_{-\infty}^\infty (u_t)^2 \rd x \rd y \rd z = \frac{1}{2}\mathfrak{m}_LA_L^2 \dot Z_L^2 +\frac{1}{2}\mathfrak{m}_R A_R^2\dot Z_R^2 + \mathfrak{m}_{LR}A_LA_R\dot Z_L\dot Z_R, \label{eq:kinetic_2QP}
\end{equation}
where the pseudomasses are given by  Eq.~\eqref{eq:pseudomass}, 
$\mathfrak{m}_L = \mathfrak{m}_R =  \frac{2\pi}{3{\beta}}$, 
and, along with the pseudomass of each QP, we have also the \emph{crossmass} (or \emph{induced mass}):
\begin{multline}
\mathfrak{m}_{LR}(w) :=   \iiint_{-\infty}^{\infty}  \frac{(z-Z_L)(z-Z_R)}{\rho_L\rho_R }  \Phi'(\rho) \Phi'(\rho_R)   \rd x \rd y \rd z  
=2\pi
 \int_{-\infty}^{\infty} \int_0^\infty\Phi'(\rho_L) \Phi'(\rho_R) \frac{\zeta (\zeta-w)}{\rho_L\rho_R } r \rd r \rd \zeta \\
=2\pi  \int_0^\infty \left\{\int_{0}^{\pi} \frac{\rho \cos^2\theta - w \cos\theta}{\rho_R} \Phi'(\rho_R) \rho \sin\theta  \rd \theta \right\}\Phi'(\rho)  \rho \rd \rho. 
\label{eq:induced_mass}
\end{multline}
The result of the integration can be found in Eq.~\eqref{eq:cross_mass_deriv}, which gives the following expression for the crossmass:
\begin{equation}
\mathfrak{m}_{LR}(w) = \frac{2\pi}{\beta  w^3} \left[ 8 \beta ^3  - e^{-{w}/{\beta }} \left(w^3 + 4 \beta  w^2 + 8 \beta ^2 w + 8 \beta ^3\right) \right].
\label{eq:cross_mass}
\end{equation}

Apparently, the fact the kinetic energy in Eq.~\eqref{eq:kinetic_2QP} contains additional terms in the case of two QPs was first noticed in \cite{Chri_CMDS6}. The term \emph{crossmass} was coined in  \cite{IvanDad_sG} for the case of interacting solitons in the case $s$GE, and the pseudo-Newtonian law of interaction of two QPs was elaborated in full detail therein as well, including a numerical solution for the trajectories. Here, we follow the line of reasoning from \cite{IvanDad_sG}.

\begin{figure}[ht]
	\centering
	\subfloat[][linear plot]{\includegraphics[width=0.49\textwidth]{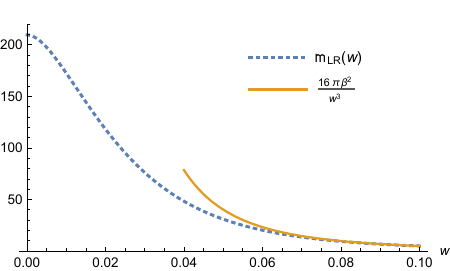}}
	\hfill
	\subfloat[][log-log plot]{\includegraphics[width=0.49\textwidth]{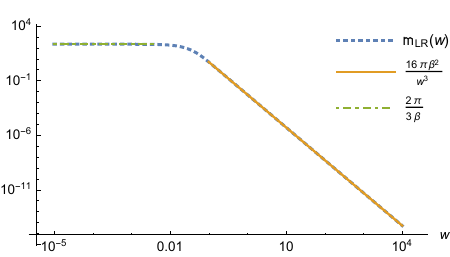}}
\caption{The crossmass $\mathfrak{m}_{LR}(w)$ between two QPs; $\beta=0.01$.}
\label{fig:crossmass}
\end{figure}

Fig.~\ref{fig:crossmass} depicts the dependence of the crossmass on the distance between the QPs for the above chosen dispersion parameter $\beta=0.01$.

\section{Law of Motion for Two Interacting QPs}

Now, we move to the essence of the coarse-grain description, namely, replacing in the original Lagrangian of the system, Eq.~\eqref{eq:Lagrange}, with the sum of  Eqs.~\eqref{eq:2QP_potential_Laplace}, \eqref{eq:2QP_potential_bihar} and \eqref{eq:kinetic_2QP}. As a result, after acknowledging Eq.~\eqref{eq:total_potent},
we obtain the following coarse-grain (discrete) Lagrangian:
\begin{equation}
\mathbb{L} =
-2\pi\frac{A_L^2+A_R^2}{\beta} -  A_L A_RV(Z_R-Z_L) +\frac{1}{2} \mathfrak{m}_L A_L^2\dot Z_L^2 +\frac{1}{2} \mathfrak{m}_R A_L^2\dot Z_R^2 + \mathfrak{m}_{LR}A_L A_R\dot Z_L\dot Z_R.
\end{equation}
Note that the first term in the above expression is constant, so only the remaining terms  contribute to the Euler--Lagrange equations with respect to the ``generalized coordinates'' $Z_L$ and $Z_R$, namely
\begin{subequations}\label{eq:pseudo_Newton}
\begin{align}
\frac{\delta\, \mathbb{L}}{\delta Z_L} &= A_L^2\frac{\rd }{\rd t}\left[ \mathfrak{m}_L \frac{\rd Z_L}{\rd t} \right] + A_LA_R \frac{\rd }{\rd t}\left[ \mathfrak{m}_{LR}\frac{\rd Z_R}{\rd t} \right] +  A_L A_R F(Z_R-Z_L) = 0 , \label{eq:pseudo_Newton1}\\
\frac{\delta\, \mathbb{L}}{\delta Z_R} &= A_LA_R \frac{\rd }{\rd t}\left[ \mathfrak{m}_{LR}\frac{\rd Z_L}{\rd t} \right]  +A_R^2\frac{\rd }{\rd t}\left[ \mathfrak{m}_L \frac{\rd Z_R}{\rd t} \right]  - A_L A_R F(Z_R-Z_L) = 0 ,  \label{eq:pseudo_Newton2}
\end{align}
\end{subequations}
where $F$ is defined in Eq.~\eqref{eq:force}.

The dynamical system in Eqs.~\eqref{eq:pseudo_Newton} for the motion of the QPs (the centers $Z_L$ and $Z_R$ of the two QPs)  differs from one governed by Newton's law of inertia due to the presence of the crossmass  terms. These terms indicate that the inertia of a QP is modified by the presence of another accelerating particle in its vicinity. As Mach put it \cite{kk73}: ``every interaction in the Universe depends on all the surrounding matter.'' For reasons that should be now obvious to the reader, we will refer to Eqs.~\eqref{eq:pseudo_Newton} as a \emph{pseudo-Newtonian} law.

As seen in Fig.~\ref{fig:crossmass}, the crossmass decays with the inverse of the cube of the distance between the QPs, which means that it becomes negligible even for QPs that are not very far from each other, and the law of acceleration is practically a Newtonian one. At closer distances, however, the crossmass terms can have a  significant impact on the law of motion. 

We solve the equations of motion \eqref{eq:pseudo_Newton} numerically, in order to understand the  mechanism of interaction between two QPs  if left to the sole action of the force acting between  them.

\section{Results and Discussion}

For definiteness we chose $\beta=0.01$. This value is not very small, and there should be no problems with very thin layers in which the force varies drastically between asymptotic scalings. 
On the other hand, $\beta=0.01$ is small enough to give a realistic picture of possible application to wave mechanics where $\beta$ is extremely small (of order of the square root of Plank's constant). 

\begin{figure}[h]
	\centering
	\subfloat[][full scale]{\includegraphics[width=0.5\textwidth]{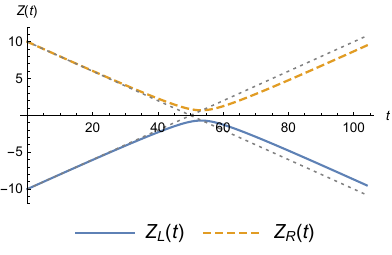}}
	\hfill
	\subfloat[][zoom]{\includegraphics[width=0.5\textwidth]{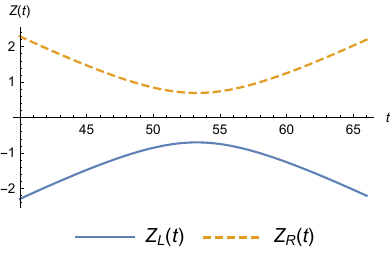}}
\caption{Trajectories of QPs for $A_L=A_R=1, \beta=0.01$ and $Z_R(0) = -Z_L(0)= 10$, $\dot Z_L(0)=-\dot Z_R(0) = 0.2$ Solid line: $Z_L(t)$; dashed line: $Z_R(t)$. In (a), thin dotted lines represent a QP's trajectories in the absence of the other QP, i.e., the lines $Z_k(0) + \dot{Z}_k(0) t$.}
\label{fig:law_motion11}
\end{figure}

\begin{figure}[h]
	\centering
	\includegraphics[width=0.5\textwidth]{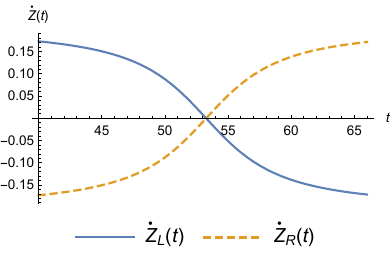}
\caption{Zoom of the velocities of the QPs for the case from Fig.~\ref{fig:law_motion11}. Solid line: $\dot{Z}_L(t)$; dashed line: $\dot{Z}_R(t)$.}
\label{eq:law_motionDeriv11}
\end{figure}

We choose the initial positions of the left and right QPs to be  $-10$ and $10$, respectively.  For such a separation, the perturbation due to the other particle at the site of the first particle will be small, hence initial speeds should be assigned for a collision to occur: $\dot Z_L(0) = -\dot Z_R(0) = 0.2$.  The first numerical experiment is for amplitudes $A_L=A_R=1$. Fig.~\ref{fig:law_motion11} shows the result form the numerical integration of Eqs.~\eqref{eq:pseudo_Newton} for the selected set of parameters.
If we find the maxima of the moduli of the QPs, they are equal to each other within the precision of the computations (see Fig.~\ref{eq:law_motionDeriv11}). So, QPs of the same masses (same amplitudes) scatter with the same velocities after the collision, save for a phase shift (compare the dashed and solid lines to the dotted ones for $t\ge 60$ in Fig.~\ref{fig:law_motion11}).

\begin{figure}[ht]
	\centering
     \subfloat[][full scale]{\includegraphics[width=0.5\textwidth]{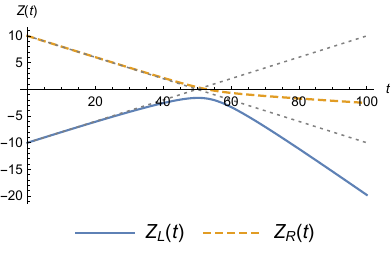}}
     \hfill
     \subfloat[][zoom]{\includegraphics[width=0.5\textwidth]{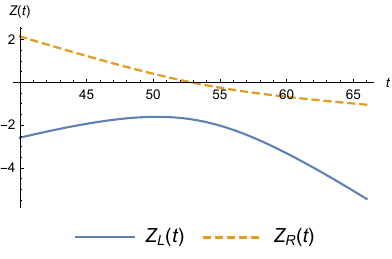}}
\caption{Trajectories of QPs for $A_L=2, A_R=1, \beta=0.01$ and $Z_R(0) = -Z_L(0)= 10$, $\dot Z_L(0)=-\dot Z_R(0) = 0.2$ Solid line: $Z_L(t)$; dashed line: $Z_R(t)$. In (a), thin dotted lines represent a QP's trajectories in the absence of the other QP, i.e., the lines $Z_k(0) + \dot{Z}_k(0) t$.}
\label{fig:law_motion12}
\end{figure}

The next numerical experiment is to consider QPs with different amplitudes. We select $A_L=1$ and $A_R=2$ (with all other parameters equal) and show the result in Fig.~\ref{fig:law_motion12}. Note that since the pseudomasses depend on the squares of the respective amplitudes, the mass of the left-going particle (initially on the right for positive abscissa)  is four times bigger than the mass of the right-going particle (initially on the left).
Now, it is clear that the more massive particle (represented by $Z_R$)  is much less affected by the force and is moving  slower than the particle $Z_L$ (the lighter one). For this reason, the two QPs do not collide at the center of the coordinate system but slightly below it.

\begin{figure}[ht]
	\centering
	\includegraphics[width=0.5\textwidth]{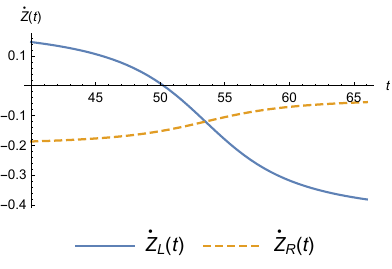}
\caption{Zoom of the velocities of the QPs for the case from Fig.~\ref{fig:law_motion12}. Solid line: $\dot{Z}_L(t)$; dashed line: $\dot{Z}_R(t)$.}
\label{eq:law_motionDeriv12}
\end{figure}

In order to assess quantitatively the magnitudes of velocities of the incoming and outgoing QPs, we found the large-$t$ values of the absolute values of the velocities. In this case, the graph of the velocities of the QPs is presented in  Fig.~\ref{eq:law_motionDeriv12}. Our numerical solution shows that, with high accuracy, both QPs attain their maximal velocity moduli at large $t$. The ratio of the two speeds after reflection is $\approx 12.1904$, which is about 3 times the expected value of $4$. Evidently, collisions of QPs of different masses are inelastic.

Finally, we would like to address the issue of the interaction  of a QP and an anti-QP. Let us set for definiteness that $A_L=1$ and $A_R=-1$.  From Eq.~\eqref{eq:force} (see also Fig.~\ref{fig:force}) it is evident that $A_LA_RF(w) <0$ for large $w$, i.e., the QP and the anti-QP repel each other. 
Imposing large initial velocity can be called ``smashing" a QP into an anti-QP.
The coupling to electromagnetic phenomena \cite{Christov2011334} has to be added to the model in order to assess its relevance to physical reality. Yet, many of the essential features of the interaction are captured by this conceptually straightforward model based on the corse-grain description. An enlightening discussion of the concept of anti-particle from the point of view of soliton theory is presented in \cite{Filippov}. 
The repulsion between matter (QPs) and antimatter (anti-QPs) means that a spontaneous annihilation can be a rare event.  When far enough apart, the QPs and anti-QPs steer clear of each other.

\section{Conclusion}

In the present paper, a new field equation is explored using the coarse-grain description (CGD), i.e., by identifying individual elements in the field called quasi-particles (QPs).

The solitary wave solutions of the new equation are shown to be pseudolocalized in the sense that the wave profile itself is not in the Sobolev space $L^2(\mathbb{R}^3)$ (square integrable functions), but its gradient and Laplacian do belong to $L^2(\mathbb{R}^3)$. It is shown that the analytical solution of the linearized equation is a very good quantitative approximation for the shape of the solitary wave, which offers the possibility to study the interaction analytically.

The pseudolocalized solutions are used as the shapes of the QPs of the new field equation, and a CGD is constructed via the so-called variational approximation (VA) in which a discrete Lagrangian is obtained for the trajectory functions of the centers of the QPs. To this end, the interaction potential is computed and shown to decay as the inverse of the distance between the QPs. Respectively, the force depends on the inverse square of the distance between the QPs at large separations. This is radically different from the currently known QPs for which the interaction potential (and the force) decay exponentially with the separation of the QPs. Thus, the first QPs that interact ``gravitationally" are presented here.

From the discrete Lagrangian, the laws of motion for two interacting QPs are derived. The notion of wave mass (\emph{pseudomass}) is introduced and an induced mass (\emph{crossmass}) is found alongside the standard mass. The interpretation is that the inertia of a QP is influenced by the presence of an accelerating QP in its vicinity, which fits well with Mach's principle. 

Similarly to the kink and anti-kink solitons of the $s$GE, the QPs of the new equation come with positive or negative amplitude. We have defined a quantity called the \emph{pseudovolume} which is the integral of the curvature of the profile. It turns out that the pseudovolume is positive when the amplitude of the respective QP is positive. The force acting between QPs is proportional to the product of their amplitudes, and hence a QP and anti-QP repulse each other if they are at moderate to large distances. At the same time, the masses of the two QPs are always positive because they are proportional to the squares of their amplitudes. An anti-QP reacts to non-gravitational forces in the same manner as a QP, but the difference shows only in the binary interaction between an anti-QP and a QP.

The coupled equations for the laws of motion were solved numerically, and the dynamics were interrogated using different amplitudes of the QPs. The scattering (not passing through each other) of QPs was confirmed, and information on the reflection was gathered. 

In summary, the present paper lays a claim that a scalar field model can be found whose QPs interact in a fashion much closer to the known properties of the interactions of real particles.

\section*{Acknowledgements} 
This work was supported in part by the National Science Fund of Ministry of Education, Science, and Youth of Republic of Bulgaria under grant DDVU02/71. The author is indebted to Dr. I. C. Christov for many discussions and numerous helpful suggestions.

\bibliographystyle{elsarticle-num}
\bibliography{pseudolocalized}


\appendix

\section{Evaluation of Integrals}

Recall that the shape of a single QP in the linear limit is given by Eq.~\eqref{eq:spherical_symmetry}.

\subsection{Potential from the Laplacian term}\label{app:laplacian}

To evaluate the integral in Eq.~\eqref{2QP_potential_Laplace2}, we can introduce again the spherical coordinates centered at $z= Z_L$. Then, $\rho_L=\rho$ and $\rho_R= \sqrt{\rho^2 - 2 \rho w \cos\theta + w^2}$. Then, the integral in Eq.~\eqref{2QP_potential_Laplace2} can be manipulated into the following
\begin{equation}
\begin{aligned}
\mathcal{I} &:= \int_0^{\infty} \!\!\! \int_0^\pi  \Phi'(\rho)\Phi'(\rho_R)   \frac{\rho^2 - \rho w \cos\theta}{\rho\rho_R} \rho^2 \sin \theta \rd \theta \rd \rho\\  
&= \int_0^{\infty} \!\!\! \int_0^\pi \Phi'(\rho)\Phi'(\rho_R) \left[ \frac{\rho^2 - w^2}{2 \rho \rho_R} + \frac{\rho^2 - 2 \rho w \cos\theta + w^2}{2 \rho \rho_R} \right]\rho^2 \sin \theta \rd \theta \rd \rho   \\
&= \int_0^{\infty} \!\!\! \int_0^\pi  \frac{1}{2}\Phi'(\rho)\Phi'(\rho_R)\ \frac{\rho^2 - w^2}{ \rho_R} \rho \sin \theta \rd \theta \rd \rho  
+ \int_0^{\infty} \!\!\! \int_0^\pi  \frac{1}{2}\Phi'(\rho)\Phi'(\rho_R) \rho_R  \rho \sin \theta \rd \theta \rd \rho  
\end{aligned}
\label{eq:2QP_potential_spherical}
\end{equation}
We observe that
\begin{equation}
\DD{\Phi(\rho_R)}{\theta} = \rho w \Phi'(\rho_R) \frac{\sin\theta}{\rho_R},
\qquad \DD{\rho_R}{\theta} = \rho w \frac{\sin\theta}{\rho_R}, \label{eq:differentials}
\end{equation}
which gives
\begin{equation} 
\mathcal{I}  =
\frac{1}{2w}\int_0^{\infty}Q_1(\rho)\Phi'(\rho)(\rho^2-w^2)\rd \rho
+\frac{1}{2}\int_0^{\infty}Q_2(\rho)\Phi'(\rho)\rd \rho,
\label{eq:2QP_potential_spherical}
\end{equation}
where the terms that depend on $\theta$ are combined in the following functions (note that $\mu := \cos\theta$):
\begin{subequations}
\begin{equation}
\begin{aligned}
Q_1(\rho) :=  \int_0^\pi w \Phi'(\rho_R) \rho \frac{\sin\theta}{\rho_R} \rd \theta 
&= \Phi(\rho_R)\big|_{\theta = \pi} -  \Phi(\rho_R)\big|_{\theta = 0}  
= \Phi(|\rho+w|) - \Phi(|\rho - w|) \\
&= \left( \frac{1-e^{-|\rho+w|/{\beta}}}{|\rho + w|} - \frac{1-e^{-|\rho-w|/{\beta}}}{|\rho - w|} \right), 
\end{aligned}
\end{equation}
\begin{equation}
\begin{aligned}
Q_2(\rho) :=  \int_0^\pi \Phi'(\rho_R)  \rho_R \rho \sin \theta \rd \theta 
&=  \int_0^\pi \left( - \frac{1 - e^{-\rho_R/\beta}}{\rho_R} + \frac{1}{\beta}e^{-\rho_R/\beta}\right) \rho\sin \theta  \rd \theta \\
&=  - \int_0^\pi \frac{1 - e^{-\rho_R/\beta}}{\rho_R} \rho\sin \theta  \rd \theta
 +\int_0^\pi \frac{1}{\beta}e^{-\rho_R/\beta} \rho\sin \theta  \rd \theta \\
&=  -  \frac{1}{w} \int_{|\rho-w|}^{|\rho+w|} [1 - e^{-\rho_R/\beta}]\rd \rho_R 
+ \frac{1}{w\beta}  \int_{-1}^{1} e^{-\frac{1}{\beta}\sqrt{\rho^2 - 2 \rho w \mu + w^2}} \rho w  \rd \mu.
\end{aligned}
\label{eq:Q2}
\end{equation}
\end{subequations}

Proceeding with the manipulations of the last two integrals, we obtain
\begin{subequations}
\begin{align}
 -  \frac{1}{w} \int_{|\rho-w|}^{|\rho+w|} [1 - e^{-\rho_R/\beta}]\rd \rho_R &= - \frac{1}{w} \left[ \rho_R + \beta e^{-\rho_R/\beta} \right]_{|\rho-w|}^{|\rho+w|} 
 = -  \frac{1}{w}\Big[ |\rho+w| - |\rho-w| + \beta\big( e^{-|\rho+w|/\beta} - e^{-|\rho-w|/\beta}\big) \Big], \\
 \frac{1}{w\beta} \int_{-1}^{1} e^{-\frac{1}{\beta}\sqrt{\rho^2 - 2 \rho w \mu + w^2}} \rho w  \rd \mu 
  &= \frac{1}{w}\left[  e^{-|\rho+w|/\beta}(\beta + |\rho +w|) - e^{-|\rho-w|/\beta}(\beta + |\rho - w|)\right] .
\end{align}
\end{subequations}
Thus, we can provide the final version for Eq.~\eqref{eq:Q2}:
\begin{equation}
Q_2(w) = \frac{1}{w}\Big[\left( e^{-|\rho+w|/\beta}-1\right)|\rho +w| -\left( e^{-|\rho-w|/\beta}-1\right)|\rho - w| \Big].
\end{equation}
Now,
\begin{multline}
\mathcal{I} =
 \frac{1}{2w}\int_0^{\infty}\left[ \frac{1-e^{-|\rho+w|/{\beta}}}{|\rho + w|}(\rho^2-w^2) - \frac{1-e^{-|\rho-w|/{\beta}}}{|\rho - w|}(\rho^2-w^2) \right] \Phi'(\rho)\rd \rho\\
- \frac{1}{2w} \int_0^{\infty} \left[ \frac{1-e^{-|\rho+w|/\beta}}{|\rho +w|}|\rho +w|^2 - \frac{1-e^{-|\rho-w|/\beta}}{|\rho - w|}|\rho - w|^2 \right] \Phi'(\rho)\rd \rho\\
= \beta^{-1}e^{-{w}/{\beta }} \left\{ -\gamma + \log \left({2 \beta }/{w}\right) + \Ei\left({w}/{\beta }\right)+e^{{2 w}/{\beta }} \left[\Ei\left(-{2 w}/{\beta }\right)-\Ei\left({w}/{\beta }\right)\right] \right\}.
\label{eq:2QP_potential_lap}
\end{multline}

\subsection{Potential from the biharmonic term}

To evaluate the integral in Eq.~\eqref{eq:2QP_potential_bihar}, applying the same notations $\rho_L=\rho$, $\rho_R = \sqrt{\rho^2 - 2\rho w \cos \theta+w^2}$, we first obtain
\begin{equation}
2\pi \beta^2 \int_{-\infty}^{\infty} \int_0^\infty \Omega(\rho)\Omega(\rho_R)  r \rd r \rd z 
= \frac{2\pi}{\beta^2} \int_{0}^{\infty} \left[\int_{0}^{\pi}\frac{e^{-\rho_R/\beta}}{\rho_R}\sin\theta \rd \theta\right]  \frac{e^{-\rho/\beta}}{\rho} \rho^2 \rd \rho .
\end{equation}
Using Eq.~\eqref{eq:differentials}, we recast the last integral as
\begin{multline}
\frac{2\pi} {\beta^2} \int_{0}^{\infty} \frac{1}{\rho w}\left[-\int_{|\rho-w|}^{|\rho+w|}e^{-\rho_R/\beta} \rd \rho_R\right]  \frac{e^{-\rho/\beta}}{\rho} \rho^2 \rd \rho 
=\frac{2\pi} {\beta^2}  \int_{0}^{\infty} \frac{\beta}{w}\left[e^{-|\rho-w|/\beta} -  e^{-|\rho+w|/\beta} \right]  e^{-\rho/\beta} \rd \rho\\ 
 = \frac{2\pi} {\beta w}  \left[ \int_{0}^{w}e^{-(w-\rho)/\beta} e^{-\rho/\beta} \rd \rho +\int_{w}^{\infty}e^{-(\rho-w)/\beta} e^{-\rho/\beta} \rd \rho -  \int_{0}^{\infty}e^{-(\rho+w)/\beta} e^{-\rho/\beta} \rd \rho \right] 
  = \frac{2\pi}{\beta} e^{-w/\beta}.  \label{eq:Vb}
\end{multline}

\subsection{Crossmass}

For the crossmass from Eq.~\eqref{eq:induced_mass}, we evaluate the following integral
\begin{equation}
\int_0^\infty\Phi'(\rho)  Q_m(\rho,w)	 \rho \rd \rho,  \quad \text{where} \quad
 Q_m(\rho,w) = \int_{0}^{\pi} \Phi'(\rho_R)  \frac{\rho \cos^2\theta - w \cos\theta}{\rho_R} \rho \sin\theta \rd \theta.
\end{equation}
Making use of the relation in Eq.~\eqref{eq:differentials}, we manipulate the inner integral from the last equation as follows
\begin{multline}
 Q_m(\rho,w)  = \frac{1}{3 \rho  w^3}  \Bigg\{ -\sgn(\rho-w) \left[\left(2 \rho ^3+w^3\right) + \left(-6 \beta ^2 \rho -3 \rho  w^2+6 \beta ^2 w\right) e^{-{|\rho - w| }/{\beta }} \right]\\
  -  \left[ \left(w^3-2 \rho ^3\right)
 +3 (\beta +\rho ) \left(2 \beta ^2+w^2+2 \beta  w\right) e^{-{|\rho +w|}/{\beta }}
 -3 \beta  e^{-{|\rho - w| }/{\beta }} \left(2 \beta ^2+w^2-2 \rho  w\right)\right] \Bigg\}.
\end{multline}
Then,
\begin{equation}
\int_0^\infty\Phi'(\rho)  Q_m(\rho,w)\rho \rd \rho  =
\frac{1}{\beta  w^3} \left[ 8 \beta ^3  - e^{-{w}/{\beta }} \left( w^3 + 4 \beta  w^2 + 8 \beta ^2 w + 8 \beta ^3\right) \right].
\label{eq:cross_mass_deriv}
\end{equation}


\end{document}